\documentclass[psfig]{aastex}

\usepackage{emulateapj5}

\def\ltsima{$\; \buildrel < \over \sim \;$}

\def\lsim{\lower.5ex\hbox{\ltsima}}
\def\gtsima{$\; \buildrel > \over \sim \;$}
\def\gsim{\lower.5ex\hbox{\gtsima}}

\begin{document}
\title{Calibrating the Galaxy Halo -- Black Hole Relation Based on the
Clustering of Quasars}

\author{J. Stuart B. Wyithe\altaffilmark{1} and Abraham
Loeb\altaffilmark{2}}

\email{swyithe@isis.ph.unimelb.edu.au; aloeb@cfa.harvard.edu}

\altaffiltext{1}{University of Melbourne, Parkville, Victoria, Australia}

\altaffiltext{2}{Harvard-Smithsonian Center for Astrophysics, 60 Garden
St., Cambridge, MA 02138}

\begin{abstract}

The observed number counts of quasars may be explained either by long-lived
activity within rare massive hosts, or by short-lived activity within
smaller, more common hosts.  It has been argued that quasar lifetimes may
therefore be inferred from their clustering length, which determines the
typical mass of the quasar host. Here we point out that the relationship
between the mass of the black-hole and the circular velocity of its host
dark-matter halo is more fundamental to the determination of the clustering
length.  In particular, the clustering length observed in the 2dF quasar
redshift survey is consistent with the galactic halo -- black-hole relation
observed in local galaxies, provided that quasars shine at $\sim
10$--$100\%$ of their Eddington luminosity. The slow evolution of the
clustering length with redshift inferred in the 2dF quasar survey favors a
black-hole mass whose redshift-independent scaling is with halo circular
velocity, rather than halo mass. These results are independent from
observations of the number counts of bright quasars which may be used to
determine the quasar lifetime and its dependence on redshift.  We show that
if quasar activity results from galaxy mergers, then the number counts of
quasars imply an episodic quasar lifetime that is set by the dynamical time
of the host galaxy rather than by the Salpeter time.  Our results imply
that as the redshift increases, the central black-holes comprise a larger
fraction of their host galaxy mass and the quasar lifetime gets shorter.

\end{abstract}

\keywords{cosmology: theory -- cosmology: observations -- quasars: general}

\section{Introduction}

The Sloan Digital Sky Survey (York et al.~2000) and the 2dF quasar redshift
survey (Croom et al.~2001a) have measured redshifts for large samples of
quasars, and determined their luminosity function over a wide range of
redshifts (Boyle et al.~2000; Fan et al.~2001a,b; Fan et al.~2003).  The
2dF survey has also been used to constrain the clustering properties of
quasars (Croom et al. 2001b). It has been suggested that quasars have
clustering statistics similar to optically selected galaxies in the local
universe, with a clustering length $R_0\sim8$Mpc. The large sample size of
the 2dF quasars also provided clues about the variation of clustering
length with redshift (Croom et al. 2001b) and apparent magnitude (Croom et
al.~2002). Quasars appear more clustered at high redshift, although with a
relatively mild trend. There is also evidence that more luminous quasars
may be more highly clustered.

As pointed out by Martini \& Weinberg~(2001) and Haiman \& Hui~(2001), the
quasar correlation length determines the typical mass of the dark matter
halo in which the quasar resides. One may therefore derive the quasar
duty-cycle by comparing the number density of quasars with the density of
host dark matter halos. The quasar lifetime then follows from the product
of the duty-cycle and the lifetime of the dark-matter halo in between major
mergers, although there is a degeneracy between the lifetime and the quasar
occupation fraction or beaming.  Preliminary results suggested quasar
lifetimes of $t_{\rm q}\sim10^6-10^7$ years, consistent with the values
determined by other methods (see Martini~2003 for a review), including the
transverse proximity effect (Jakobsen et al.~2003) and counting arguments
relative to the local population of remnant supermassive black holes
(SMBHs; see Yu \& Tremaine~2002). Kauffmann \& Haehnelt~(2002) used a
detailed semi-analytic model (Kauffmann \& Haehnelt~2000) to predict the
correlation length of quasars and its evolution with redshift. They found
that their model reproduces the correct correlation length as well as its
redshift evolution for present-day quasar lifetimes of $t_{\rm q}\sim10^7$
years.

In this work we argue that the correlation length of quasars is
fundamentally determined by the relation between the masses of SMBHs and
their host galactic halo rather than by the quasar lifetime. 
We find that the amplitude of the correlation length depends on the product
of the SMBH--halo relation and the typical fraction of the Eddington
luminosity at which quasars shine. Moreover, we show that the evolution of
the correlation length is sensitive to how the SMBH -- halo relation
evolves with redshift, and therefore to the physics of SMBH formation and
quasar evolution.

The paper is organized as follows. In \S~\ref{cf} and \S~\ref{MbhMhalo} we
discuss the calculation of the correlation function of quasars with a
particular apparent magnitude $B$, and the different scenarios for the
evolution of the SMBH -- galaxy halo relation.  In \S~\ref{cfobs} we
compare fiducial model correlation functions to the results of the 2dF quasar
redshift survey (Croom et al.~2001b,c).  We find that the shallow evolution
of the correlation length implies that SMBH mass has a redshift-independent
scaling with the circular velocity of the host dark matter halo rather than
with its mass. The ranges of the normalizations in
the SMBH -- galaxy halo relation and of the fraction of Eddington allowed
by observations of the observed quasar correlation function are explored in
\S~\ref{epseta}. In \S~\ref{lf} we examine the quasar lifetime by requiring
that observational constraints from both the correlation and luminosity
functions be satisfied simultaneously. Finally, in \S~\ref{disc} we summarize our
results and discuss a very simple, physically motivated model that
satisfies all constraints with no free parameters. Throughout the paper we
adopt the set of cosmological parameters determined by the {\em Wilkinson
Microwave Anisotropy Probe} (WMAP, Spergel et al. 2003), namely mass
density parameters of $\Omega_{m}=0.27$ in matter, $\Omega_{b}=0.044$ in
baryons, $\Omega_\Lambda=0.73$ in a cosmological constant, and a Hubble
constant of $H_0=71~{\rm km\,s^{-1}\,Mpc^{-1}}$.

\section{The Correlation Function of Quasars}
\label{cf}

The mass correlation function between halos of mass $M_1$ and $M_2$,
separated by a co-moving distance $R$ is (see Scannapieco \& Barkana~2003 and
references therein)
\begin{eqnarray}
\nonumber
\xi_{\rm m}(M_1,M_2,R)&=& \frac{1}{2\pi^2}\int dk k^2 P(k) \\
&\times&\frac{\sin(kR)}{kR}W(kR_1)W(kR_2),
\end{eqnarray}
where 
\begin{equation}
R_{1,2} = \left(\frac{3 M_{\rm 1,2}}{4\pi\rho_{\rm m}}\right)^{1/3},
\end{equation}
$W$ is the window function (top-hat in real space), $P(k)$ the power
spectrum and $\rho_{\rm m}$ is the cosmic mass density.  The dark-matter
halo correlation function for halos of mass $M$ is obtained from the
product of the mass correlation function $\xi_{\rm m}(M,M,R)$ and the
square of the ratio between the variances of the halo and mass
distributions.  This ratio, $b$, is defined as the halo bias; its value
for a halo mass $M$ may be approximated using the Press-Schechter formalism
(Mo \& White~1996), modified to include non-spherical collapse (Sheth, Mo
\& Tormen~2001)
\begin{eqnarray}
\label{bias}
\nonumber
b(M,z) = 1 + \frac{1}{\delta_{\rm c,0}}&&\left[\nu^{\prime2}+b\nu^{\prime2(1-c)}\right.\\
&&\left.-\frac{\nu^{\prime2c}/\sqrt{a}}{\nu^{\prime2c}+b(1-c)(1-c/2)}\right],
\end{eqnarray}
where $\delta_{\rm c,0}$ ($\approx 1.69$) is the critical overdensity
threshold for spherical collapse, $\delta_{\rm c}(z)=\delta_{\rm c,0}/D(z)$,
$D(z)$ is the growth factor at redshift $z$, $\sigma$ is the variance on a 
mass-scale $M$, $\nu\equiv {\delta_{\rm c}^2(z)}/{\sigma^2(M)}$, 
$\nu^\prime\equiv\sqrt{a}\nu$, $a=0.707$, $b=0.5$ and $c=0.6$.
This expression yields an accurate approximation to the halo bias determined from
N-body simulations (Sheth, Mo \& Tormen~2001).

We need the halo correlation function between halos of different masses. 
The bias (equation~\ref{bias}) yields
the excess probability relative to the mass distribution of finding a halo at any point.
We may therefore determine the halo correlation function for halos of mass $M_1$ and $M_2$
using the product of the biases for individual masses, 
\begin{equation}
\label{cchalo}
\xi_{\rm h}(M_1,M_2,R) = b(M_1,z)b(M_2,z)\xi_{\rm m}(M_1,M_2,R)D(z)^2.
\end{equation}
If the correlation function between halos forming at different redshifts
is required, then the more general  analytic treatment of
Scannapieco \& Barkana~(2003) should be used.

We would like to constrain the relationship between SMBH mass, $M_{\rm
bh}$, and galactic halo mass, $M_{\rm halo}$, by constructing a theoretical
quasar correlation function for comparison with observational
data. Observations of the quasar correlation function measure the
clustering of quasars of a certain luminosity, rather than the clustering
of halos or SMBHs of a particular mass. We therefore need to associate the
$B$-band luminosity $L_{\rm B}$ of a quasar with the halo mass of its host
galaxy $M_{\rm halo}$. We begin by assuming quasars to have a spectral
energy distribution corresponding to the median of their population (Elvis
et al.~1994). We then allow the SMBH to shine at a fraction $\eta$ of its
Eddington luminosity, so that in solar units, the $B$-band luminosity of
the quasar is
\begin{equation}
\frac{L_{\rm B}}{L_{\rm B,\odot}}=5.73\times10^{12} \,\eta 
\left(\frac{M_{\rm bh}}{10^9M_\odot}\right).
\end{equation}
We also need to specify a relation between SMBH and halo mass [$M_{\rm
bh}=f(M_{\rm halo})$, see \S~\ref{MbhMhalo}]. Given a luminosity $L_{\rm
B}$, quasar pairs shining with Eddington fractions $\eta_1$ and $\eta_2$
have their correlation function specified by equation~(\ref{cchalo}) where
$M_{1,2}=f^{-1}(M_{\rm bh,1,2})$ and $M_{\rm bh,1,2}=L_{\rm
B}/(5.73\times10^3 \,\eta_{1,2}L_{\rm B,\odot}) M_\odot$.

The correlation function for a population of quasars with a luminosity
$L_{\rm B}$ can therefore be computed by drawing pairs of $\eta$-values
from an appropriate probability distribution, $dP_{\rm obs}/d\eta$. The
correlation function of quasars with luminosity $L_{\rm B}$ is
\begin{equation}
\xi_{\rm q}(L_{\rm B}) = \langle \xi_{\rm h}(L_{\rm B},\eta_1,\eta_2) \rangle,
\end{equation}
where $\eta_{1,2}$ are drawn from $dP_{\rm obs}/d\eta$ and angular brackets
denote an average over the probability distribution of $\eta$-values.

We consider two distributions for $\eta$ in this paper. In the first case 
all quasars shine at their Eddington luminosity, so that
$dP_{\rm obs}/d\eta\propto\delta(\eta-1)$, where $\delta$ is the Dirac delta
function.  This corresponds to an idealized quasar lightcurve that is a
tophat-function in time, $\eta (t)=\Theta(-t)\Theta(t-t_{\rm q})$, whereby
the quasar shines at the Eddington luminosity ($\eta=1$) throughout its
lifetime, $t_{\rm q}$.  In the second case, the observed
distribution is assumed be flat in the logarithm of $\eta$ so that
$dP_{\rm obs}/d(\log\eta)$ is constant. This is the form of distribution
expected for a quasar lightcurve of the exponential form
$\Theta(-t)\eta(t)=\exp{(-{t}/{t_{\rm q}})}$.

\section{the M$_{\rm bh}$-M$_{\rm halo}$ relation}
\label{MbhMhalo}

Dormant SMBHs are ubiquitous in local galaxies (Magorrian et al.~1998). The
masses of these SMBHs scale with physical properties of their hosts
(e.g. Magorrian et al., 1998; Merritt \& Ferrarese 2001; Tremaine et
al. 2002).  Motivated by local observations (Ferrarese~2002), we assume a
relation $M_{\rm bh}=f\left(M_{\rm halo},z\right)$ between SMBH mass and
the mass of the host dark matter halo. This relation may be applied to the
calculations of the correlation function of quasars because the masses of
SMBHs scale the same way with the physical properties of their host
galaxies in both quiescent and active galaxies (McLure \& Dunlop~2002). The
function $f$ is constrained by local observations (Ferrarese~2002), and
we consider two different forms for its redshift dependence:

\begin{itemize}

\item {\bf Case A}: we assume that SMBH mass is correlated with the halo
circular velocity. This scenario is supported empirically by Shields et
al.~(2003) who studied quasars out to $z\sim3$ and demonstrated that the
relation between $M_{\rm bh}$ and the stellar velocity dispersion does not
evolve with redshift. This is expected if the mass of the black-hole is
determined by the depth of the gravitational potential well in which it
resides, as would be the case if growth is regulated by feedback from
quasar outflows (e.g. Silk \& Rees~1998; Wyithe \& Loeb~2003).  Expressing
the halo circular velocity, $v_c$, in terms of the halo mass, $M$, and
redshift, $z$, the redshift dependent relation between the SMBH and halo
masses may be written as
\begin{eqnarray}
\label{eps}
\nonumber M_{\rm bh}(M_{\rm halo},z) &=&\mbox{const} \times v_c^5\\
&&\hspace{-25mm}= \epsilon M_{\rm halo} \left(\frac{M_{\rm
halo}}{10^{12}M_{\odot}}\right)^{\frac{2}{3}}
[\zeta(z)]^\frac{5}{6}(1+z)^\frac{5}{2},
\end{eqnarray}
where $\epsilon$ is a constant, $\zeta(z)$ is close to unity and
defined as $\zeta\equiv [(\Omega_m/\Omega_m^z)(\Delta_c/18\pi^2)]$,
$\Omega_m^z \equiv [1+(\Omega_\Lambda/\Omega_m)(1+z)^{-3}]^{-1}$,
$\Delta_c=18\pi^2+82d-39d^2$, and $d=\Omega_m^z-1$ (see equations~22--25 in
Barkana \& Loeb 2001 for more details). 

\item {\bf Case B}: we consider the scenario in which the SMBH mass maintains
the same dependence on its host halo mass at all redshifts, namely
\begin{eqnarray}
\label{eps2}
\nonumber M_{\rm bh}(M_{\rm halo},z) &=&\mbox{const} \times M_{\rm halo}^{5/3}\\
&&\hspace{-25mm}= \epsilon M_{\rm halo} \left(\frac{M_{\rm
halo}}{10^{12}M_{\odot}}\right)^{\frac{2}{3}}
[\zeta(0)]^\frac{5}{6}.
\end{eqnarray}
\end{itemize}

\noindent The normalizing constant in these relations has an
observed\footnote{ We have used the normalization derived by
Ferrarese~(2002) under the simplifying assumption
that the virial velocity of the halo
represents its circular velocity.} value of $\epsilon=\epsilon_{\rm
SIS}\approx 10^{-5.1}$ at $z=0$ (Ferrarese~2002); where
SIS denotes the underlying
assumption that the halo mass profile resembles a Singular Isothermal Sphere.  In the following section we
construct model correlation functions assuming both of these scenarios, and
compare the results to observations. We will show that case {\bf A} is
consistent with the data of the 2dF quasar redshift survey, while case
{\bf B} is not.

\subsection{Summary of models}

Before proceeding to the comparison with the observed correlation length
we label six models of interest:

\begin{itemize}

\item {\bf (AI)} The fiducial model: $M_{\rm bh}\propto v_{c}^5$, $\eta=1$,
$\epsilon=\epsilon_{\rm SIS}$.

\item {\bf (AII)} $M_{\rm bh}\propto v_{c}^5$, $dP_{\rm
obs}/d(\log{\eta})=\mbox{const}$, $\epsilon=\epsilon_{\rm SIS}$.

\item {\bf (AIII)} $M_{\rm bh}\propto v_{c}^5$, $dP_{\rm
obs}/d(\log{\eta})=\mbox{const}$, \\$\epsilon=10\epsilon_{\rm SIS}$.

\item {\bf (BI)} $M_{\rm bh}\propto M^{3/2}$, $\eta=1$,
$\epsilon=\epsilon_{\rm SIS}$.

\item {\bf (BII)} $M_{\rm bh}\propto M^{3/2}$, $dP_{\rm
obs}/d(\log{\eta})=\mbox{const}$, \\$\epsilon=\epsilon_{\rm SIS}$.

\item {\bf (BIII)} $M_{\rm bh}\propto M^{3/2}$, $dP_{\rm
obs}/d(\log{\eta})=\mbox{const}$, \\$\epsilon=10\epsilon_{\rm SIS}$.

\end{itemize}

\noindent In models {\bf AII, AIII, BII} and {\bf BIII}, the range
$0.01\le\eta\le1$ is considered.  We refer to the above labels when
presenting our results through the remainder of the paper.

The models described above utilize a $M_{\rm bh}$--$M_{\rm halo}$
relation with a normalization derived under the simplifying assumption of a flat rotation
curve. However, the circular velocity at the radii probed by
observations of galaxies is likely to be larger than the circular
velocity at their virial radius. The normalization
$\epsilon$ in equations~(\ref{eps}) and (\ref{eps2}) may
therefore be larger than 
the value $\epsilon_{\rm SIS}$ considered
here (Ferrarese~2002). For any universal halo profile, this difference is degenerate with a simple
renormalization of $\eta$.  The results in the following section may be applied to a
case where the normalization is $\epsilon'=C\epsilon$
by substituting a correspondingly lower value of the Eddington luminosity
$\eta'=\eta/C$. The allowed range of the product $\epsilon\eta$ is explored
in \S~\ref{epseta}.

\begin{figure*}[t]
\epsscale{1.5}  \plotone{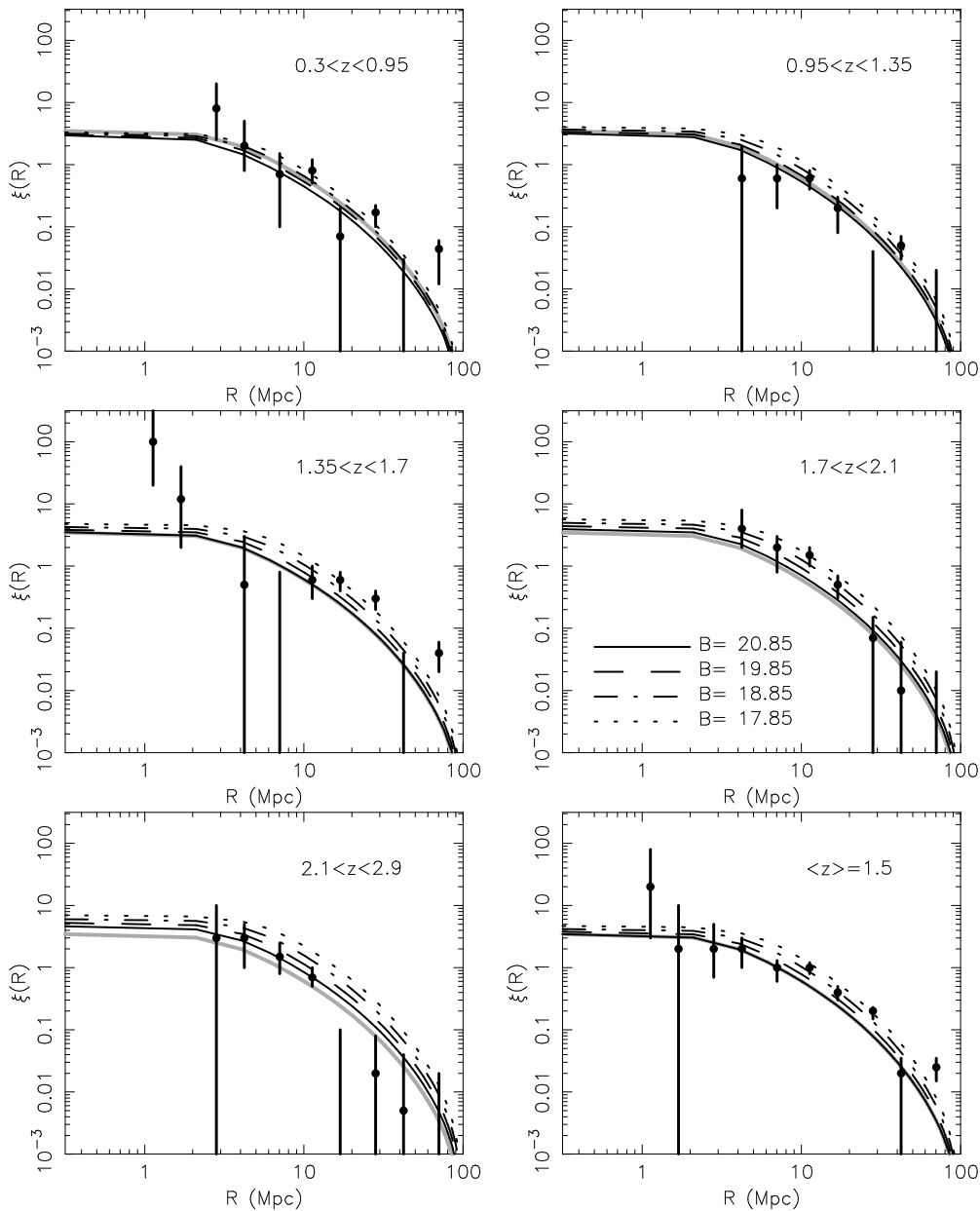}
\caption{\label{fig1}Predicted correlation function at various redshifts,
in comparison to the 2dF data (Croom et al.~2001b). The dark lines show
the correlation function predictions for quasars of various apparent B-band
magnitudes ($B$). The 2dF limit is $B\sim20.85$. The lower right panel
shows data from entire 2dF sample in comparison to the theoretical
prediction at the mean quasar redshift of $\langle z\rangle=1.5$. The
$B=20.85$ prediction at this redshift is also shown by thick gray lines in the
other panels to guide the eye. This case assumes $M_{\rm bh}\propto v_c^5$,
normalized to local observations, and $\eta=1$ (case {\bf AI}).}
\end{figure*}

\section{Comparison to Observations}
\label{cfobs}

\begin{figure*}[t]
\epsscale{1.5}  \plotone{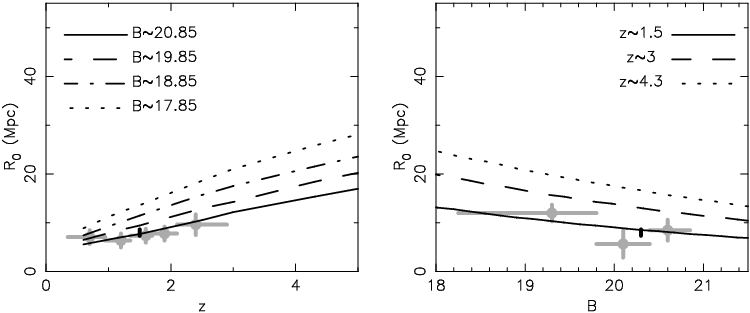}
\epsscale{1.5}  \plotone{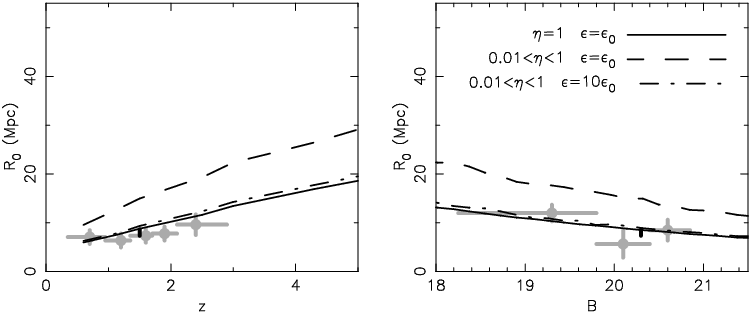}
\caption{\label{fig2}The correlation length, $R_{0}$ [defined through
$\xi_{\rm q}(R_0)=1$] plotted as a function of redshift (left hand panel)
and apparent magnitude (right hand panel). The data is from the 2dF quasar
redshift survey (Croom et al.~2001b,c), with the gray error bars showing 
the correlation length from different sub-samples, and the dark error bar
at $z=1.5$ (left hand panels) and $B=20.3$ (right hand panels) showing 
the correlation length for the full sample. The theoretical model assumes
$M_{\rm bh}\propto v_c^5$. In the upper panels $\eta=1$ (case {\bf AI}),
and the curves are shown for different apparent magnitudes (left), and
different redshifts (right). In the lower panels we show curves
corresponding to different choices for $\eta$ and $\epsilon$, (cases {\bf
AII} and {\bf AIII}) assuming $B=20.3$ and $z=1.5$ respectively. }
\end{figure*}

\begin{figure*}[t]
\epsscale{1.5}  \plotone{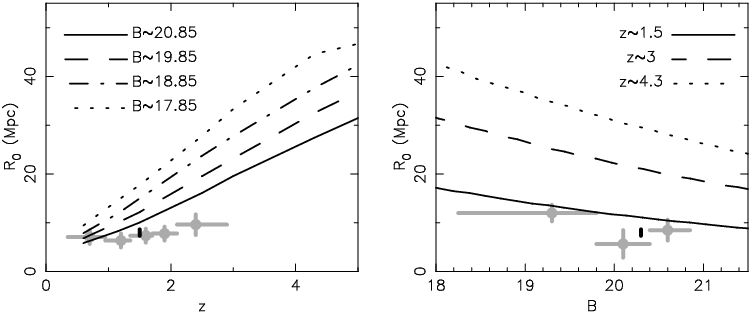}
\epsscale{1.5}  \plotone{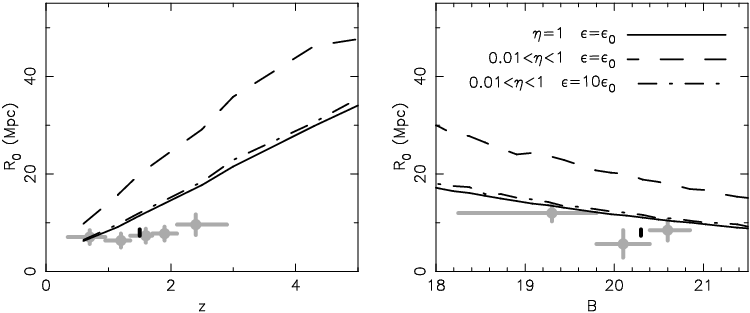}
\caption{\label{fig3}The correlation length, $R_{0}$ as a function of
redshift (left hand panel) and apparent B-band magnitude (right hand
panel). The data is from the 2dF QRS (Croom eta al.~2001b,c), 
with the gray error bars showing 
the correlation length from different sub-samples, and the dark error bars
at $z=1.5$ (left hand panels) and $B=20.3$ (right hand panels) 
showing the correlation length for the full sample.
The model assumes $M_{\rm bh}\propto M_{\rm halo}^{5/3}$. In the upper panels
$\eta=1$ (case {\bf BI}), and the curves are shown for different apparent
magnitudes (left), and different redshifts (right). In the lower panels we
show curves corresponding to different choices for $\eta$ and $\epsilon$
(cases {\bf BII} and {\bf BIII}), assuming $B=20.3$ and $z=1.5$ respectively. }
\end{figure*}

Next we compare model correlation functions to observations from the 2dF
quasar redshift survey (Croom et al.~2001b,c).  Figure~\ref{fig1} shows the
correlation function obtained by assuming that SMBH mass scales with halo
circular velocity and that all quasars shine at their Eddington luminosity
(case~{\bf AI}).  We show the predicted correlation function at various
redshifts, and several apparent B-band magnitudes\footnote{We refer to 
the Johnson $B$-magnitude, which for a typical quasar is related to the
photographic b-band used in the 2dF quasar redshift survey by $B\approx b+0.06$
(Goldschmidt \& Miller~1998)}, as determined from the
relation
\begin{eqnarray}
\label{absm}
\nonumber B &=& 5.5 -2.5\log_{10}\left(\frac{L_{\rm B}}{L_{{\rm B},\odot}}\right) +
5\log_{10}\left(\frac{D_{\rm L}}{10\mbox{pc}}\right)\\
&-&2.5(1+\alpha_\nu)\log_{10}\left(1+z\right),
\end{eqnarray}
where $\alpha_\nu=-0.5$ is the typical slope of the quasar power-law
continuum (with a frequency dependence of the spectral flux $f_\nu\propto
\nu^{\alpha_\nu}$), and $D_{\rm L}$ is the luminosity distance. The
magnitude limit of the 2dF quasar redshift survey is $B\sim20.85$, and so
the data should be compared with the model correlation function
corresponding to $B\sim20-21$ (solid and dashed lines). The lower right panel of
Figure~\ref{fig1} shows the 2dF correlation function generated from the
entire quasar sample. This data is compared with a model correlation
function computed at the median quasar redshift of the survey ($\langle
z\rangle\sim1.5$). The $B=20.85$ prediction at this redshift is replotted in
the other panels of figure~\ref{fig1} to guide the eye (thick gray curves).
We find that the correlation function depends more sensitively on redshift
than on luminosity over the ranges under consideration.

\subsection{The correlation function for models with $M_{\rm bh}\propto v_{\rm c}^5$}

Variation of clustering with redshift and apparent magnitude is demonstrated more
clearly in Figure~\ref{fig2}. The upper panels show the dependences of the
clustering length, $R_{0}$ [defined through the condition $\xi_{\rm
q}(R_0)=1$] for the fiducial case where the SMBH mass scales with halo
circular velocity and quasars shine at their Eddington limit (case {\bf
AI}). The correlation length is plotted for various values of apparent
magnitude as a function of redshift (left panel), and various redshifts as
a function of apparent magnitude (right panel).  The data for the
correlation length as a function of redshift was computed
from a sample having $B\la20.85$, drawn from the 2dF quasar redshift survey
(Croom et al.~2001b,c). The gray error bars show the values for $R_0$ in
different redshift bins, while the correlation length measured for the
whole sample (plotted at $z=1.5$ and $B=20.3$ in the left and right hand
panels respectively) is shown by the dark error bar.  We find the
predictions for $R_0$ to be consistent with observations.  In particular,
both the value of the clustering length and its slow evolution with
redshift and apparent magnitude are consistent with current data. There is
little variation of the clustering length with apparent magnitude over the
range considered, although there is a tendency for brighter quasars to be
more highly clustered, as was noted by Croom et al.~(2002).

Note that the K-correction assumed in equation~(\ref{absm}) is only valid
out to $z\sim3$ where Ly$\alpha$ absorption enters into the $B$-band. As a
result the extrapolations of redshift dependence to $z\ga3$ shown in
Figures~\ref{fig2} and \ref{fig3} do not include Ly$\alpha$ absorption in
the calculation of the quasar luminosity. However, these extrapolations are
included to illustrate the evolution of the clustering length at yet higher
redshifts (as measured in bands redward of the redshifted Ly$\alpha$ line).

The lower panels of Figure~\ref{fig2} show the corresponding dependence of
the clustering length on redshift and apparent magnitude for quasars that
shine at Eddington fractions $\eta$ lower than unity.  Only the cases of
$B=20.3$, corresponding approximately to the median quasar B-magnitude 
(left hand panel) and $z=1.5$ (right hand panel) are shown, and
the curve corresponding to $\eta=1$ (case {\bf AI}) is reproduced from the
upper panels for comparison (solid lines). Assuming the local normalization
$\epsilon=\epsilon_{\rm SIS}$ and that $\eta$ ranges between 0.01 and 1
with a constant $dP_{\rm obs}/d(\log\eta)$ (case {\bf AII}), the typical
quasar of apparent magnitude $B$ must be powered by a more massive SMBH and
hence reside in a more massive galaxy than in the $\eta=1$ case. This
results in a larger clustering length for quasars of a given apparent
magnitude $B$.  The resulting evolution, indicated by the dashed lines,
appears to be incompatible with the data.  The increase in $R_0$ may be
compensated for by an increase in the normalization of the $M_{\rm
BH}-M_{\rm halo}$ relation.  The dot-dashed curves show the result of
increasing this normalization relative to local observations by a factor of 10
($\epsilon=10\epsilon_{\rm SIS}$). This factor equals the inverse of the
median of the distribution $dP/d(\log{\eta})=\mbox{const}$. 
The resulting curve (corresponding to case {\bf AIII}) is very similar
to our fiducial case ($\eta=1$).  

These results demonstrate that while a spread in the values of $\eta$ for
different quasars results in the correlation function being measured over a
wider range of halo masses, the correlation length is primarily sensitive
to the characteristic halo mass, as specified by the characteristic value
of $\eta$. Furthermore, the results demonstrate the important point that
the correlation length measures the product of the normalization in the
$M_{\rm bh}$--$M_{\rm halo}$ relation and the median fraction of the
Eddington luminosity at which quasars shine, $\eta_{\rm med}$. With respect
to the correlation function, this quantity $\epsilon\eta_{\rm med}$ (which
is the same in cases {\bf AI} and {\bf AIII}) is more fundamental to the
correlation function than the quasar lifetime since it predicts the
correlation length independent from consideration of quasar number
counts. It is important to note that since the normalization of the local
$M_{\rm BH}-M_{\rm halo}$ relation is estimated ($\epsilon=\epsilon_{\rm
SIS}$, Ferrarese~2002), 
the value of the quasar correlation length at low
redshift implies that quasars spend most of
their bright phase shining near their Eddington luminosity.

\subsection{The correlation function for models with 
$M_{\rm bh}\propto M_{\rm halo}^{5/3}$}

We next examine whether the correlation function observed by the 2dF survey
may be used to constrain the evolution of the $M_{\rm bh}$--$M_{\rm halo}$
relation.  In the previous sub-section, we have considered the case of a
SMBH mass that scales with the fifth power of circular velocity independent
of redshift (equation~\ref{eps}, case {\bf A}).  The alternative case~{\bf
B} example postulates that the SMBH mass has a redshift independent scaling
with the halo mass rather than halo circular velocity
(equation~\ref{eps2}).  The upper panels of Figure~\ref{fig3} plot the
correlation length in this case with $\eta=1$ and $\epsilon=\epsilon_{\rm
SIS}$ (case {\bf BI}). 
Both the evolution of the correlation length with
redshift for different apparent magnitudes (left hand panel), and the
variation of correlation length with a given apparent magnitude at various
redshifts (right hand panel) are shown.  The case {\bf B} $M_{\rm
bh}-M_{\rm halo}$ relation results in a correlation length that varies
significantly more with redshift than the data requires. The lower panels
show examples with values of $\eta$ that differ from unity.  Only the cases
of $B=20.3$ (left hand panel) and $z=1.5$ (right hand panel) are shown,
and the result assuming $\eta=1$ (case {\bf BI}) is reproduced from the
upper panels for comparison (solid lines). If $\epsilon=\epsilon_{\rm
SIS}$, and $\eta$ ranges between 0.01 and 1, with $dP_{\rm
obs}/d\eta\propto\mbox{const}$ (case {\bf BII}), then the typical quasar of
apparent magnitude $B$ must have a larger clustering length (dashed lines).
This may be compensated for by increasing the normalization of the $M_{\rm
BH}$--$M_{\rm halo}$ relation by a factor of 10 ($\epsilon=10\epsilon_{\rm
SIS}$, case {\bf BIII}) as considered before (dot-dashed lines).

The rapid evolution with redshift that results from a case {\bf B} $M_{\rm
BH}$--$M_{\rm halo}$ relation could be diminished if the luminosities of
quasars were a higher fraction of their Eddington limit at higher
redshift. However we have seen that the locally observed normalization of
the $M_{\rm bh}$--$M_{\rm halo}$ relation requires most present-day quasars
to be shining near their Eddington luminosity. Quasars at high redshift
would therefore have to be super-Eddington, shining at $\sim\eta\sim(1+z)^{5/2}$ 
($\eta\sim30$ at $z\sim3$) in order to reproduce the observed slow evolution 
of the clustering length. This requirement on $\eta$ may be somewhat relaxed 
in models having $\epsilon>\epsilon_{\rm SIS}$, as is discussed in the next 
section.

\section{Constraining $\eta$ and $\epsilon$ using the quasar correlation function}
\label{epseta}

\begin{figure*}[t]
\epsscale{.9}  \plotone{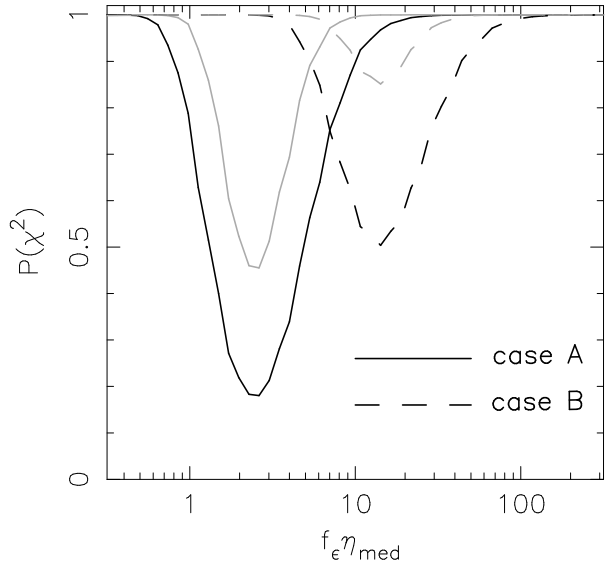}
\caption{\label{fig4} The probability $P(\chi^2)$ of case~{\bf A} and case~{\bf B}
models as a function of $f_\epsilon\eta_{\rm med}$. The dark lines correspond
to fits to the redshift dependent correlation length $R_0(z)$ measured by the
2dF quasar redshift survey (Croom et al.~2001b). The gray lines show 
results for a hypothetical larger future sample whose errors on $R_0$ are
reduced by a factor of $\sqrt{2}$ relative the the 2dF quasar survey.}
\end{figure*}

The $M_{\rm bh}$--$M_{\rm halo}$ relation for local galaxies requires a
choice for the relationship between the circular velocity at the radius
probed by observations of galaxies, and the circular velocity at the virial 
radius. This relationship depends on the dark-matter mass profile and is
still uncertain due to the gravitational interaction between the baryons
(which cool and often dominate gravity near the center of galaxies) and the
dark matter.  Ferrarese~(2002) presented $M_{\rm bh}$--$M_{\rm halo}$
relations for three different cases, which may be summarized in the present
notation by $\epsilon=\epsilon_{\rm SIS}$, $\epsilon=\epsilon_{\rm
NFW}=3.7\epsilon_{\rm SIS}$, and $\epsilon=\epsilon_{\rm
S02}=25\epsilon_{\rm SIS}$.  The second case corresponds to the universal
profile of Navarro, Frenk \& White~(1996, NFW), and the third to the
normalization derived based on the weak lensing study of Seljak~(2002).

The results in the preceding sections have been presented for fiducial
models with $(\epsilon,\eta)=(\epsilon_{\rm SIS},1)$, ($\epsilon_{\rm
SIS},0.1$) and ($10\epsilon_{\rm SIS},0.1$). It was demonstrated that the
correlation function is primarily dependent on the product between
$\epsilon$ and the median value of $\eta$, namely $\epsilon\eta_{\rm
med}$. This raises the possibility that $\epsilon\eta_{\rm med}$ could be
constrained using correlation function data. We have computed the evolution
of the correlation length $R_0(z)$ for a set of models having a range of
values for the product $\epsilon\eta_{\rm med}\equiv
f_\epsilon\epsilon_{\rm SIS}\eta_{\rm med}$, and calculated the reduced
$\chi^2$-statistics as well as the confidence with which a model is
excluded by the 2dF data, $P(\chi^2)$. In figure~\ref{fig4} we
plot the resulting $P(\chi^2)$ as a function of $f_\epsilon\eta_{\rm med}$
(dark lines).

The results suggest that Case~{\bf A} models (solid lines) favor $f_\epsilon\eta_{\rm
med}\sim1$--$5$ indicating that $\epsilon$ is a factor of a few larger than
$\epsilon_{\rm SIS}$ for $\eta \sim 1$, as would be expected from the
universal NFW profile, or that $\eta_{\rm med}$ has a maximum value of a
few tenths.  If $\epsilon=\epsilon_{\rm S02}$, case~{\bf A} requires
$\eta_{\rm med}\sim0.08$.  Figure~\ref{fig4} shows that Case~{\bf B} models
(dashed lines) are more restricted, requiring $f_\epsilon\eta_{\rm med}\sim10-30$. This
implies that $\epsilon=\epsilon_{\rm SIS}$ and $\epsilon=\epsilon_{\rm
NFW}$ cannot be accommodated since the required values of $\eta_{\rm
med}\ga10$ and $\eta_{\rm med}\ga3$ respectively, are considered unphysical
in most accretion models (Nityananda \& Narayan 1982; see, however,
Ruszkowski \& Begelman 2003).  The more extreme case of
$\epsilon=\epsilon_{\rm S02}$ may be marginally accommodated by case~{\bf
B}; however even in this case the quasars must shine somewhat 
in excess of their Eddington limit.

Figure~\ref{fig4} also shows $P(\chi^2)$ as a function of
$f_\epsilon\eta_{\rm med}$ for a hypothetical future survey that would
yield the same evolution of $R_0$ with redshift, but with errors that are
smaller by a factor of $\sqrt{2}$ (light lines).  Future surveys will
better constrain the gradient $dR_0/dz$, and help discriminate further
between different models of the $M_{\rm bh}$--$M_{\rm halo}$ relation.

\begin{figure*}[t]
\epsscale{1.9}  \plotone{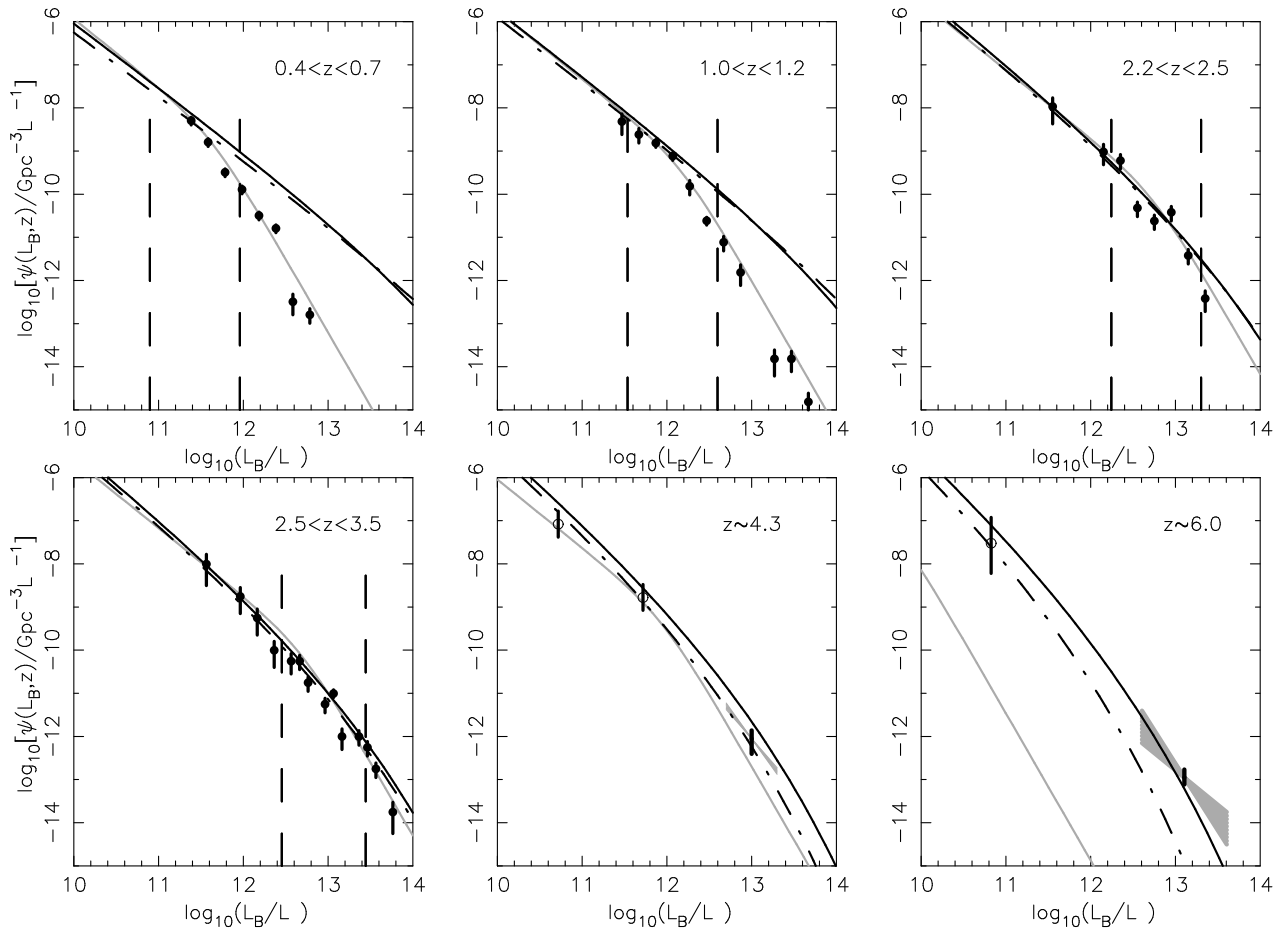}
\caption{\label{fig5}Comparison of the model and observed quasar luminosity
functions at various redshifts. The data points at $z\la4$ are summarized
in Pei~(1995), while the light lines show the double power-law fit to the
2dF quasar luminosity function (Boyle et al.~2000).  At $z\sim4.3$ and
$z\sim6.0$ the data is from Fan et al.~(2001a;2001b;2003). The gray regions
show the 1-$\sigma$ range of logarithmic slope ($[-2.25,-2.75]$ at
$z\sim4.3$ and $[-1.6,-3.1]$ at $z\sim6$), and the vertical bars show the
uncertainty in the normalization. The open circles show data points
converted from the X-ray luminosity function (Barger et al.~2003) of low
luminosity quasars using the median quasar spectral energy distribution
(Elvis et al.~1994). The model luminosity functions shown correspond to
cases {\bf AI} (solid curves) and a model assuming case~{\bf B} evolution 
with $\epsilon =\epsilon_{\rm S02}$, $\eta=1$ and a constant lifetime 
of $t_{\rm q}=10^7$ years (dot-dashed curves). The
dashed pairs of vertical lines show the range corresponding to apparent
B-magnitudes between 18.25 and 20.85.}
\end{figure*}

Our discussion so far has been independent of the quasar
luminosity function. In the next section we discuss how the luminosity
function may be used to determine the redshift evolution of the quasar
lifetime.

\section{The Luminosity Function and Quasar Lifetime}
\label{lf}

In the previous sections we have demonstrated that the evolution of the
$M_{\rm bh}$--$M_{\rm halo}$ relation with redshift may be constrained
directly from the quasar correlation function. We have shown that the
critical parameter determining the SMBH mass is the halo circular velocity
rather than the halo mass. We reiterate that this result does not rely on
the luminosity function of quasars. On the other hand, the number counts of
quasars are proportional to the quasar lifetime.  Therefore, the quasar
lifetime and its evolution with redshift may be determined by comparing the
observed quasar number counts to the SMBH mass function. The latter may be
obtained from the Press-Schechter~(1976) mass-function of dark-matter halos
(with the improvement of Sheth \& Tormen~1999) combined with the $M_{\rm
bh}$--$M_{\rm halo}$ relation.

Theoretical models of the quasar luminosity function often associate quasar
activity with major mergers (e.g. Kauffmann \& Haehnelt~2000; Wyithe \&
Loeb~2003). A simple model that attributes quasar activity to major
mergers, assumes a $M_{\rm bh}$--$M_{\rm halo}$ relation that is described
by equation~(\ref{eps}), and sets the quasar lifetime equal to the
dynamical time of a galactic disk, accurately reproduces the entire optical
and X-ray luminosity functions of quasars at redshifts between 2 and 6
(Wyithe \& Loeb~2003). At lower redshifts ($z\la2$), this prescription
correctly predicts the luminosity function at luminosities below the
characteristic break (Boyle et al.~2000). 
In difference to Wyithe \& Loeb~(2003) we have assumed the quasar activity
to take place in the larger SMBH before coalescence of the SMBH binary formed 
during the merger (Volonteri et al.~2002). Figure~\ref{fig5} 
shows the comparison of this model with observations. 
Specifically, we show the fiducial case {\bf AI} (which corresponds to the
model presented in Wyithe \& Loeb~2003) using equation~(\ref{eps}) with
$\epsilon=\epsilon_{\rm SIS}$ and $\eta=1$ (solid curves). We have shown that
case~{\bf AI} describes the observed quasar correlation function (see 
figures~\ref{fig1}-\ref{fig2}). However as discussed in \S~\ref{epseta}, 
case-{\bf B} models for the $M_{\rm bh}$--$M_{\rm halo}$ relation also provide 
acceptable fits to $R_0(z)$ so long as $\epsilon\ga10\epsilon_{\rm SIS}$ and 
$\eta\sim1$. Therefore a corresponding luminosity function model was also
generated for this case assuming a constant lifetime of $t_{\rm q}=10^7$ years. 
We have plotted this model in figure~\ref{fig5} (dot-dashed lines). 

We find that our fiducial model (case {\bf AI}; solid lines) with a quasar
lifetime that equals the dynamical time of a galactic disk, reproduces the
observed luminosity function over a wide range of redshifts and
luminosities (see Wyithe \& Loeb~2003). Our second model 
(case {\bf B}, $\epsilon\ga10\epsilon_{\rm SIS}$, $\eta=1$, $t_{\rm q}=10^7$; 
dot-dashed lines) also reproduces the observed luminosity function at $z\la4.5$.
However at $z\sim6$ this model underestimates the observed number counts by 
two orders of magnitude. Case~{\bf B} evolution for the
 $M_{\rm bh}$--$M_{\rm halo}$ relation is therefore challenged by
the existence of the highest redshift quasars, which would imply
a quasar lifetime that exceeds the age of the universe at $z\sim6$. Moreover,
given the Salpeter~(1964) time of $\sim4\times10^7$ years, the SMBHs powering
these quasars would grow to 25 times their initial mass during their lifetime.
Provided that quasar activity is trigged by major mergers, the 
requirement that the $M_{\rm bh}$--$M_{\rm
halo}$ relation reproduce the observed evolution in quasar clustering,
together with the observed evolution in the number counts of quasars
therefore implies case {\bf A} evolution, and a quasar lifetime that is set 
by the dynamical time of the host galaxy. 
We reinforce the important qualitative result that since the
evolution of quasar correlations require SMBHs to take a larger fraction of
the halo mass at high redshift (case {\bf A}, see equation~\ref{eps}), the observed
evolution in the number counts requires a quasar lifetime that is shorter
at higher redshifts, scaling approximately as the dynamical time of a dark
matter halo [$\propto(1+z)^{-3/2}$].

We point out that the model luminosity functions fail to reproduce the
rapid drop in the density of luminous quasars seen at low redshift. This
drop is thought to be due to the consumption of cold gas in galaxies by
star formation (e.g. Kauffmann \& Haehnelt~2000) 
and the inhibition of accretion onto massive halos at late
times (e.g. Scannapieco \& Oh~2004). However, most of the quasars
in the 2dF quasar redshift survey have $B$-magnitudes that lie
below the characteristic break in the luminosity function at low 
redshift (Boyle et al.~2000; Croom et al.~2001). 
This can be seen in Figure~\ref{fig5} 
where the vertical dashed lines show the luminosities bracketing the
apparent magnitude range $18.25<B<20.85$. Since the model correctly predicts 
the luminosity function within most of this luminosity range at all redshifts, 
we are justified in deriving combined constraints from the luminosity
and correlation functions of quasars.

\section{Do bright low-redshift quasars shine near their Eddington Luminosity?}

In \S~\ref{cfobs} we demonstrated that the 2dF quasar sample has a
correlation length that is consistent with the local $M_{\rm bh}-M_{\rm
halo}$ relation, and that quasars shine near their Eddington luminosity.  
In addition it was shown that this consistency extends to
both bright and faint subsamples of quasars. The brightest 2dF quasars
($18.1\leq B\leq19.9$) presented by Croom et al.~(2002) are mostly brighter
than the characteristic break luminosity in the differential quasar number
counts (e.g. Boyle et al.~2000). These bright quasars have a lower
abundance than expected from merging halos at $z\la 2$ (see
figure~\ref{fig5}).
Either the lifetime (or occupation fraction) of bright quasars is lower
than expected at $z\la 2$ or else these quasars shine well below their
Eddington limit.  Figure~\ref{fig2} indicates that the slow dependence of
$R_0$ with $B$ reported by Croom et al.~(2002) is consistent with the
expected trend if all quasars shine near their Eddington luminosity. This
consistency suggests that bright quasars are under represented at $z\la 2$
because of a reduced quasar lifetime (or a reduced occupation fraction)
rather than a reduced value of $\eta$ during their brightest phase.

\section{discussion}
\label{disc}

In this paper we have demonstrated that the relationship between the mass
of a SMBH and its host dark-matter halo is {\it the} fundamental quantity
that determines the clustering of quasars. In particular the $M_{\rm
bh}$--$M_{\rm halo}$ relation and its evolution with redshift, may be
determined directly from the evolution in the correlation length of
quasars, independent of consideration of the quasar luminosity function or
assumptions about the quasar lifetime.  Beginning with the locally observed
$M_{\rm bh}$--$M_{\rm halo}$ relation, we have shown that the observed
correlation length of quasars in the 2dF quasar redshift survey is
consistent with SMBHs that shine at $\sim 10$--$100\%$ of their Eddington
luminosity during their bright quasar episode. Moreover, the evolution of
the clustering length with redshift is consistent with a SMBH mass whose
redshift-independent scaling is with the circular velocity of the host dark
matter halo (see equation~\ref{eps}). In contrast, it appears that a
relation for the SMBH mass whose redshift-independent scaling is with the
mass of the host dark matter halo (equation~\ref{eps2}) would have resulted
in too much evolution of the clustering length with redshift.

The portion of the 2dF quasar survey from which the data used in this study
were drawn, contains $\sim10^4$ quasars. Upon completion, the Sloan Digital
Sky Survey will have spectra for ten times this number of quasars, spread
over a larger redshift range (York et al.~2000). This will allow more
accurate determination of the correlation length and its evolution, and
hence provide more accurate constraints the SMBH population. The selection
of quasars for the spectroscopic survey of SDSS at $z<3$ is restricted to
$i^*<19$. Thus, most of the low redshift SDSS quasars will fall at luminosities
that are brighter than the characteristic break in the luminosity function
(e.g. Boyle et al.~2000). These bright quasars have a lower abundance than
expected from the number of halos merging at $z\la 2$. The reason for this low 
abundance may be either that the lifetime (or occupation fraction) of bright quasars is 
lower than expected at $z\la 2$, or alternatively that the fraction of the Eddington 
limit at which the quasars shine is small. The 2dF quasar sample straddles
the luminosity function break. We have shown that the similarity in the
clustering statistics of sub-samples of quasars having different ranges in
luminosity suggest that the lifetime (or occupation fraction) of bright quasars is 
lower than expected at $z\la 2$, but that these quasars also shine near their
Eddington luminosity. Measurements of the correlation length 
as a function of quasar luminosity in SDSS can help further distinguish 
between these two possibilities.

As we have shown in a previous paper (Wyithe \& Loeb~2003),
feedback--regulated growth of SMBHs during active quasar phases implies
$M_{\rm bh}\propto v_c^5$, consistent with the inferred relation for
galactic halos in the local universe (Ferrarese 2002). Assuming $\eta=1$ as
suggested by observations of both low and high-redshift quasars
(Floyd~2003; Willott, McLure \& Jarvis~2003), we have shown that only
$\sim7\%$ of the Eddington luminosity output needs to be deposited into the
surrounding gas in order to unbind it from the host galaxy over the
dynamical time of the surrounding galactic disk. The power-law index of 5
in the $M_{\rm bh}$--$v_c$ relation, is larger than the values of 4--4.5
inferred from the local relation between $M_{\rm bh}$ and the stellar
velocity dispersion $\sigma_\star$ (Merritt \& Ferrarese 2001; Tremaine et
al. 2002); the difference originating from the observation that the
$v_c$--$\sigma_\star$ relation is shallower than linear for stellar bulges
embedded in cold dark matter halos (Ferrarese~2002). Feedback--regulated
growth also implies that the $M_{\rm bh}$--$v_{\rm c}$ relation should be
independent of redshift, as implied by the slow evolution in the
clustering length of quasars.

A very simple model can therefore be constructed to describe the
correlation length of quasars and its evolution in the 2dF survey, as well
as the number counts of quasars out to very high redshift ($z\sim 6$). The
model includes four simple assumptions, but no free parameters: {\it (i)}
the locally observed $M_{\rm bh}$--$M_{\rm halo}$ relation extends to high
redshifts through equation~(\ref{eps}) with the SMBH mass scaling as halo
circular velocity to the 5-th power; {\it (ii)} SMBHs shine near their
Eddington luminosity with a universal spectrum (Elvis et al.~1994) during
luminous quasar episodes; (iii) quasar episodes are associated with major
galaxy mergers; and (iv) the quasar lifetime is set by the dynamical time
of the host galaxy.

The assumption of a SMBH mass that scales with halo circular velocity
independently of redshift is supported by the observations of Shields et
al.~(2003) that there is no evolution in the $M_{\rm bh}$--$\sigma_\star$
relation out to $z\sim3$. The assumption that quasars shine near their
Eddington limit is supported by observations of high and low redshift
quasars (Floyd~2003; Willott, McLure \& Jarvis~2003).  Overall, the sample
of available data on the clustering properties and the number counts of
quasars is most readily explained by a quasar lifetime that is set by the
dynamical time of the host galaxy rather than by the Salpeter (1964)
$e$-folding time for the growth of its mass.

\acknowledgements 

This work was supported in part by NASA grant NAG 5-13292, and by NSF
grants AST-0071019, AST-0204514 (for A.L.). JSBW wishes to thank Princeton
University Observatory for their hospitality during the course of this
work.

\end{document}